\def\year{2020}\relax
\documentclass[letterpaper]{article} % DO NOT CHANGE THIS
\usepackage{aaai20}  % DO NOT CHANGE THIS
\usepackage{times}  % DO NOT CHANGE THIS
\usepackage{helvet} % DO NOT CHANGE THIS
\usepackage{courier}  % DO NOT CHANGE THIS
\usepackage[hyphens]{url}  % DO NOT CHANGE THIS
\usepackage{graphicx} % DO NOT CHANGE THIS
\usepackage{graphicx}  % DO NOT CHANGE THIS
\usepackage[table,xcdraw]{xcolor}
\usepackage{multirow}

\usepackage{natbib}
\usepackage{amsmath}
\usepackage{algorithm}
\usepackage{algpseudocode}

\usepackage[belowskip=0pt,aboveskip=0pt]{caption}
\usepackage{xcolor}

\usepackage{booktabs}
\usepackage{arydshln}
\usepackage{array}

\title{On Analyzing Annotation Consistency in Online Abusive Behavior  Datasets}
\author{Md Rabiul Awal,\textsuperscript{1} Rui Cao,\textsuperscript{2} Roy Ka-Wei Lee,\textsuperscript{1} Sandra Mitrović\textsuperscript{3}\\ % All authors must be in the same font size and format. Use \Large and \textbf to achieve this result when breaking a line
\textsuperscript{\rm 1}University of Saskatchewan, Canada\\
\textsuperscript{\rm 2}University of Electronic Science and Technology of China, China \\
\textsuperscript{\rm 3}Istituto Dalle Molle di Studi sull'Intelligenza Artificiale, Switzerland \\
mda219@mail.usask.ca, caorui0503@gmail.com, roylee@cs.usask.ca, sandra.mitrovic@idsia.ch % email address must be in roman text type, not monospace or sans serif
}

\begin{document}

\maketitle

\begin{abstract}
Online abusive behavior is an important issue that breaks the cohesiveness of online social communities and even raises public safety concerns in our societies. Motivated by this rising issue, researchers have proposed, collected, and annotated online abusive content datasets. These datasets play a critical role in facilitating the research on online hate speech and abusive behaviors. However, the annotation of such datasets is a difficult task; it is often contentious on what should be the true label of a given text as the semantic difference of the labels may be blurred (e.g., abusive and hate) and often subjective. In this study, we proposed an analytical framework to study the annotation consistency in online hate and abusive content datasets. We applied our proposed framework to evaluate the consistency of the annotation in three popular datasets that are widely used in online hate speech and abusive behavior studies. We found that there is still a substantial amount of annotation inconsistency in the existing datasets, particularly when the labels are semantically similar. %The proposed framework is also robust and can be easily applied to evaluate future datasets.
\end{abstract}

\section{Introduction}
Misbehavior in online social media such as cyberbullying, propagation of hate speeches, and abusive content have become an increasing problem. Such online misbehavior has not only sowed discord among individuals or communities online but also resulted in violent hate crimes \cite{Williams19,relia2019race,mathew2019spread}. Therefore, it is a pressing issue to detect and curb such misbehavior in online social media.

Traditional machine learning and deep learning approaches have been proposed to detect online misbehavior automatically. The recent surveys \citep{schmidt2017survey,fortuna2018survey} have comprehensively summarized these methods. Most of the automatic online misbehavior detection methods are supervised text classification methods trained and tested on annotated datasets. As such, the quality of the annotation has direct implications on detection algorithms' performance and the insights gained from the online misbehavior research studies. 

Three popular datasets are widely used in online misbehavior studies: \textbf{WZ} \citep{waseem2016hateful,waseem2016you}, \textbf{DT} \citep{DavidsonWMW17}, and the recently published \textbf{FOUNTA} \cite{FountaDCLBSVSK18} dataset. \cite{waseem2016hateful} first collected and annotated the \textbf{WZ} Twitter dataset into four classes: racism, sexism, both, and neither. \cite{waseem2016hateful} subsequently enhanced the dataset by controlling the bias introduced by annotators. \cite{DavidsonWMW17} argued that hate speech should be differentiated from offensive tweets; some tweets may contain hateful words but should be labeled as offensive as they did not meet the threshold of classifying them as hate speech. The researchers collected the \textbf{DT} dataset and manually annotated the dataset into three categories: offensive, hate, and neither. In a recent study, \cite{FountaDCLBSVSK18} proposed the \textbf{FOUNTA} dataset. This dataset went through two rounds of annotations. In the first round, annotators are required to classify tweets into three categories: normal, spam, and inappropriate. Subsequently, the annotators were asked to refine further the labels of the tweets in the ``inappropriate'' category. Specifically, the final version of the dataset consists of four classes: normal, spam, hate,  and abusive.

While these datasets have facilitated many online misbehavior studies, few analyses have been done to evaluate and benchmark the quality of these datasets. The annotation of online misbehavior datasets is a challenging tasks. Firstly, the difference between certain labels may be subtle \citep{DavidsonWMW17,FountaDCLBSVSK18}. Secondly, the manual annotation process is often subjected to the annotator's biasness \citep{waseem2016you}. Therefore, we proposed an analytical framework to examine the annotation consistency in online misbehavior datasets. Included in our proposed framework is a two-step pipeline, which enables us to identify potential mislabeling and contentious annotation in the datasets.

\begin{figure*}[t]
    \centering
	\includegraphics[width=0.9\textwidth]{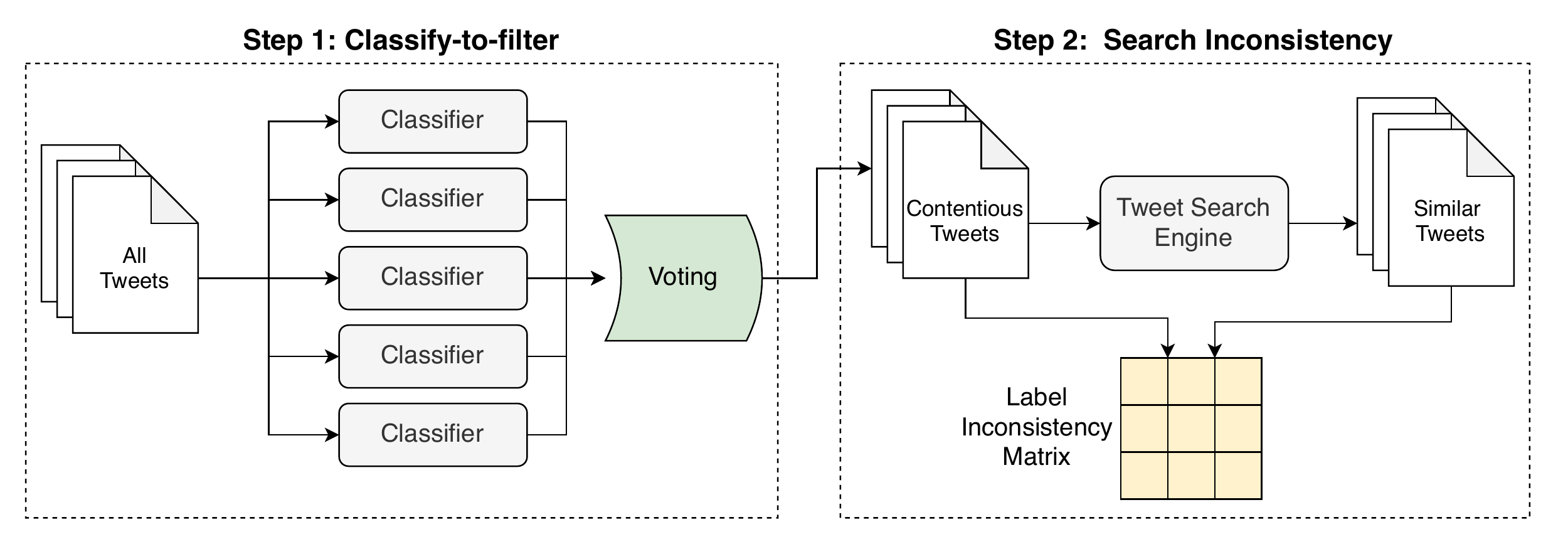}
	\caption{Overall annotation consistency analysis framework}
	\label{fig:framework}
\end{figure*}

We summarize our main contributions as follows:
\begin{itemize}
    \item We proposed a novel analytical framework to examine the annotation consistency in the online misbehavior dataset.
    \item We applied our proposed framework to analyze three popular real-world and publicly available datasets. To the best of our knowledge, we are the first study that quantitatively and qualitatively compares existing online misbehavior datasets. 
    \item Our analysis showed that there is a substantial amount of annotation inconsistency in the existing datasets. We also empirically demonstrate case studies where the annotation inconsistency is likely to occur in the datasets. 
\end{itemize}

\section{Annotation Consistency Analysis Framework}
Figure \ref{fig:framework} shows our proposed annotation consistency analysis framework. Included in the analytical framework is a two-step process. In the first step, we train a set of classifiers to predict the labels of a given dataset of tweets. Voting will then be performed to vote for contentious tweets, i.e., tweets that are wrongly classified by more than half of the classifiers. The intuition is that it is more challenging to classify tweets that are annotated with contentious labels. 
For example, in Table \ref{tbl:example}, the tweet \textit{t1} is identified as contentious when more than half of the classifiers mispredicted its label. A potential reason for the wrong classification may be due to \textit{t1}, which is labeled as \textit{Hate}, sharing very similar attributes with other tweets that are labeled as \textit{Offensive}. Such contentious labeling is likely to confuse the classifier, resulting in the wrong prediction. In the second step, the set of retrieved contentious tweets are used as input queries into a search engine to find similar tweets in the dataset. Finally, we construct an annotation inconsistency matrix by comparing the labels of the contentious tweets and the retrieved similar tweets. The underlying assumption is that potential inconsistencies arise when the labels of the contentious tweet and its similar tweet are different. For example, in Table \ref{tbl:example}, the search engines return \textit{t2} as the most similar tweet to the contentious tweet \textit{t1}. When we compare the label of \textit{t1} and \textit{t2}, we notice that the two tweets have different labels, flagging a potential annotation inconsistency for the tweet \textit{t1}.
%
%
% \begin{table}[htp]
% \centering
% \caption{Tweets example}
% \label{tbl:example}
% \begin{tabular}{c|c|c|c}
% \hline 
% \textbf{Id} & \textbf{Tweet}       & \textbf{Label} & \textbf{\begin{tabular}[c]{@{}c@{}} Contentious\end{tabular}} \\ \hline \hline
% t1 & You are such a b*tch & Hate & Yes \\ \hline
% t2 & Don't be such a b*tch & Offensive & No \\ \hline
% t3 & B*tch please, try hard! & Offensive & No \\
% \hline
% \end{tabular}
% \end{table}

\subsection{Step 1: Classify-to-filter}
The \textit{classify-to-filter} step can be further broken down into two stages: \textit{classification} and \textit{voting}.

In the \textit{classification} stage, we adopt an ensemble approach to train five different text classifiers on a given online misbehavior dataset. The commonly-used traditional machine learning and deep learning models are selected for our text classification task. 
\begin{table}[h!]
\centering
\caption{Tweets example}
\label{tbl:example}
\begin{tabular}{cccc}
% \hline
\toprule
\textbf{Id} & \textbf{Tweet}       & \textbf{Label} & \textbf{\begin{tabular}[c]{@{}c@{}} Contentious\end{tabular}} \\ \midrule
t1 & You are such a b*tch & Hate & Yes \\ 
t2 & Don't be such a b*tch & Offensive & No \\ 
t3 & B*tch please, try hard! & Offensive & No \\ 
\bottomrule
\end{tabular}
\end{table}
Specifically, we use Logistic Regression (LR), Naive Bayes (NB), Single-layer Convolutional Neural Network (CNN), Recurrent Neural Network (RNN), and Convolutional Long-Short Term Memory network (C-LSTM) as the classifiers in this step. For LR and NB, we trained these classifiers using the tweets' word-level term frequency-inverse document frequency (tf-idf) features. For the deep learning models, we use pre-trained GloVe word embeddings \citep{PenningtonSM14} to represent the words in the tweets, which are subsequently used as input for the classifiers. Each classifier is trained using 5-fold cross-validation, and the predictions on the tweets in the validation set are recorded for voting. 

In the \textit{voting} stage, we consolidate the predictions made by five classifiers and identify the contentious tweets. Specifically, given a tweet, if three or more classifiers predicted its label wrongly, we would place this tweet into the \textit{contentious tweets} set. While the incorrect prediction may be attributed to inconsistency in annotation, there could also be other reasons. For example, a tweet may contain rare words, and there are insufficient data to train the models well to classify this tweet. Therefore, we perform another step to further verify whether it is annotation inconsistency that led to incorrect predictions. 

\subsection{Step 2: Search Inconsistency}
In this step, we utilize the retrieved set of contentious tweets as input into our search engine to retrieve similar tweets. Specifically, given a query contentious tweet, $t_q$, the search engine aims to retrieve its most similar tweet, $t_s$, from the dataset. To measure the similarity between tweets, we compute the cosine similarity between the tweets' tf-idf representation. The cosine similarity between two tweets are computed as follows:  

\begin{figure*}[t] 
	\centering
	\setlength{\tabcolsep}{0pt} % Default value: 6pt
	\renewcommand{\arraystretch}{0} % Default value: 1
	\begin{tabular}{ccc}
		\includegraphics[scale = 0.55]{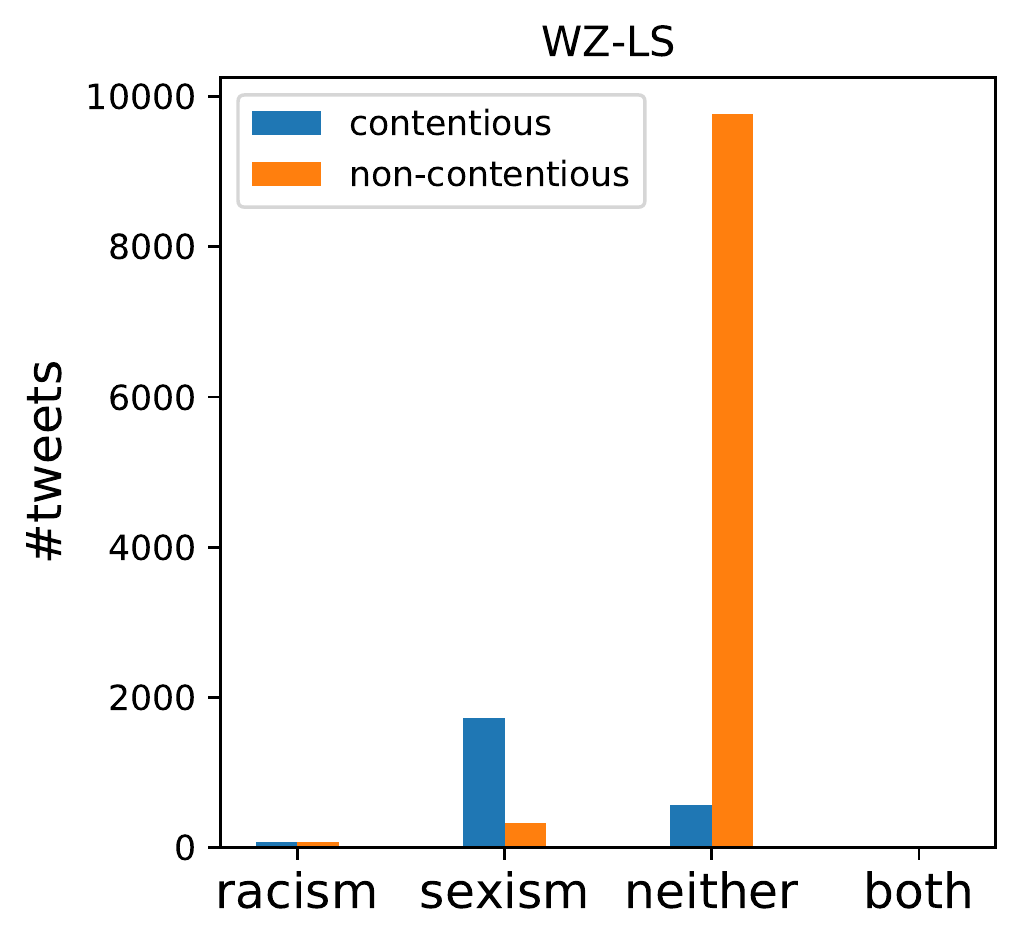} &  
        \includegraphics[scale = 0.55]{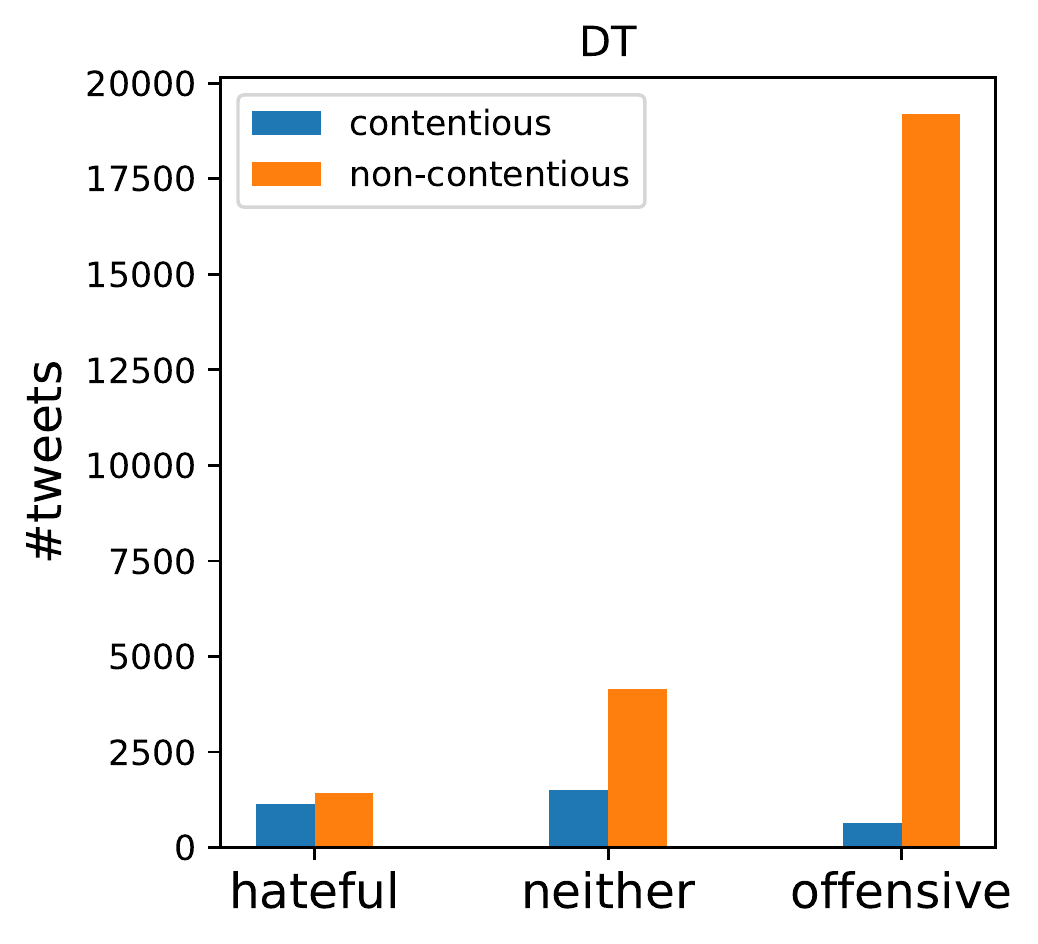} & 
        \includegraphics[scale = 0.55]{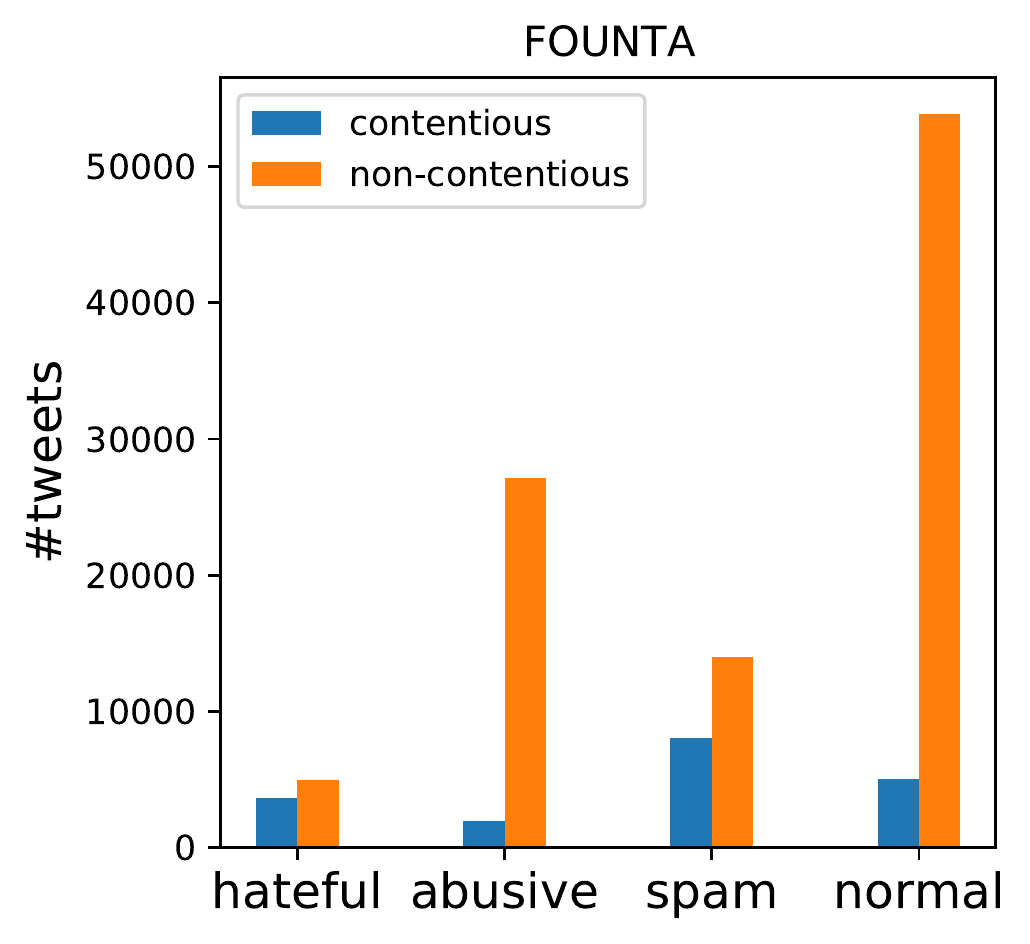}
	\end{tabular}
	\caption{Breakdown distributions of contentious and non-contentious tweets from \textbf{WZ} (left), \textbf{DT} (middle), and \textbf{FOUNTA} (right) retrieved in step 1 of the annotation consistency analysis framework. }
	\label{fig:stage_1_results}
\end{figure*}

\begin{equation}
\begin{aligned}
    cos\_sim(t_q,t_s) = \frac{\sum_{w \in t_q \cap t_s} t_q^w t_s^w}{\sqrt{\sum_{w\in t_q}{t_q^w}^2} * \sqrt{\sum_{w\in t_s}{t_s^w}^2}}
\end{aligned}
\end{equation}

\noindent where $t_q^w$ is the tf-idf weight of term $w$ in the query tweet $t_q$, and $t_s^w$ is the tf-idf weight of term $w$ in the similar tweet $t_s$. We compute the cosine similarity between each query tweet $t_q$ and all tweets in the dataset, i.e., $t_s \in T$, and select the tweet with the highest cosine similarity score as the similar tweet to the query tweet. 

Finally, we compare the annotated labels of the $t_q$ and $t_s$: if the two annotated labels %are different
disagree, we flag out that $t_q$ might have an annotation inconsistency as the (a) classifiers find it hard to classify this tweet, and (b) its annotated label is different from its most similar tweet. The annotation inconsistencies in the contentious tweets are subsequently reported in annotation inconsistency matrices in the next step.  

\section{Evaluation and Discussion}
We applied our proposed annotation consistency analysis framework on the three popular datasets, which are widely used in online misbehavior studies: \textbf{WZ} \citep{waseem2016hateful,waseem2016you}, \textbf{DT} \citep{DavidsonWMW17}, and \textbf{FOUNTA} \cite{FountaDCLBSVSK18}. The summary statistics of the datasets are presented in Table \ref{tab:dataset}. Note that we combined the number the tweets in \citep{waseem2016hateful,waseem2016you} to form the current \textbf{WZ} dataset. %However, as only the the ids of the tweets are released, we retrieve the text of the tweets using Twitter's APIs and some of the tweets have been deleted by Twitter due to their inappropriate content. Thus, our dataset is slightly smaller than the original dataset reported in \cite{ParkF17}.

\begin{table}[htbp]
\centering
  \caption{Summary statistics of datasets}
  \label{tab:dataset}
  \begin{tabular}{ccm{12em}}
    \toprule
    \textbf{Dataset} & \textbf{\#tweets} & \textbf{Classes (\#tweets)}\\ \midrule
    WZ & 13,202 & racism (82), sexism (3,332), both (21), neither (9,767)\\ \hdashline 
    DT & 24,783 & hate (1,430), offensive (19,190), neither (4163)\\ \hdashline
    FOUNTA & 99,999 & normal (53,851), abusive (27,150), spam (14,029), hate (4,965) \\ \bottomrule
\end{tabular}%}
\end{table}
%
% \begin{table}[htbp]
% \centering
%   \caption{Summary statistics of datasets}
%   \label{tab:dataset}
%   \begin{tabular}{c|c|p{12em}}
%     \hline
%     \textbf{Dataset} & \textbf{\#tweets} & \textbf{Classes (\#tweets)}\\\hline\hline
%     WZ & 13,202 & racism (82), sexism (3,332), both (21), neither (9,767)\\\hline
%     DT & 24,783 & hate (1,430), offensive (19,190), neither (4163)\\\hline
%     FOUNTA & 99,999 & normal (53,851), abusive (27,150), spam (14,029), hate (4,965) \\\hline
% \end{tabular}%}
% \end{table}
%
%
Figure \ref{fig:stage_1_results} shows the breakdown distributions of contentious and non-contentious tweets retrieved in step 1 of our proposed analytical framework. We observe that contentious tweets are found from all labels in the three datasets, i.e., the five classifiers made mistakes in predicting the true label of all kinds of tweets. Specifically, in \textbf{WZ}, the classifiers have incorrectly predicted most of the sexism tweets. In \textbf{DT}, almost half of the hateful tweets are wrongly predicted. Similar observations are made in \textbf{FOUNTA}, with hateful and spam tweets seeing a higher percentage of misclassification. 

\begin{table}[h!]
% \centering
\caption{Annotation Inconsistency matrix for WZ}
\label{tab:wz_matrix}
\begin{tabular}{cccccc}
\cmidrule[\heavyrulewidth]{3-6}
\multicolumn{1}{l}{} & \multicolumn{1}{l}{} & \multicolumn{4}{c}{\cellcolor[HTML]{EFEFEF}\textbf{Contentious Tweet Label}} \\ \cmidrule{3-6} 
\multicolumn{1}{l}{} &  & \cellcolor[HTML]{EFEFEF}Racism & \cellcolor[HTML]{EFEFEF}Sexism & \cellcolor[HTML]{EFEFEF}Both & \cellcolor[HTML]{EFEFEF}Neither \\ \midrule
\multicolumn{1}{c}{\cellcolor[HTML]{EFEFEF}} & \cellcolor[HTML]{EFEFEF}Racism & 16 & 0 & 0 & 1 \\ %\cline{2-6} 
\multicolumn{1}{c}{\cellcolor[HTML]{EFEFEF}} & \cellcolor[HTML]{EFEFEF}Sexism & 9 & 662 & 10 & 222 \\ %\cline{2-6} 
\multicolumn{1}{c}{\cellcolor[HTML]{EFEFEF}} & \cellcolor[HTML]{EFEFEF}Both & 0 & 4 & 0 & 1 \\ %\cline{2-6} 
\multicolumn{1}{c}{\multirow{-4}{*}{\cellcolor[HTML]{EFEFEF}\textbf{\begin{tabular}[c]{@{}c@{}}Similar \\Tweet \\ Label\end{tabular}}}} & \cellcolor[HTML]{EFEFEF}Neither & 26 & 754 & 5 & 218 \\ \bottomrule
\end{tabular}
\end{table}

As discussed earlier in the section, there could be multiple reasons for the misclassification. For instance, the hate speech detection problem may be hard as the tweets within the same label have high variance, or there might be insufficient training data. In this paper, we are interested to understand how much of the misclassification can be attributed to annotation inconsistency. Table \ref{tab:wz_matrix}, \ref{tab:dt_matrix}, and \ref{tab:founta_matrix} shows the annotation inconsistency matrix generated in step 2 of our analytical framework for \textbf{WZ}, \textbf{DT}, and \textbf{FOUNTA} respectively.

%\begin{figure}[h]
%    \centering
%	\includegraphics[width=0.3\textwidth]{filtered_wz.png}
%	\caption{WZ-LS - contentious vs non-contentious tweets}
%	\label{fig:inconsistency}
%\end{figure}
%
%\begin{figure}[h]
%    \centering
%	\includegraphics[width=0.3\textwidth]{filtered_dt.png}
%	\caption{DT - contentious vs non-contentious tweets}
%	\label{fig:inconsistency}
%\end{figure}

%\begin{figure}[h]
%    \centering
%	\includegraphics[width=0.3\textwidth]{filtered_founta.png}
%	\caption{FOUNTA - contentious vs non-contentious tweets}
%	\label{fig:inconsistency}
%\end{figure}
%
%
%
%
% \begin{table}[h]
% %\centering
% \caption{Annotation Inconsistency matrix for WZ}
% \label{tab:wz_matrix}
% \begin{tabular}{cc|c|c|c|c|}
% \cline{3-6}
% \multicolumn{1}{l}{} & \multicolumn{1}{l|}{} & \multicolumn{4}{c|}{\cellcolor[HTML]{EFEFEF}\textbf{Contentious Tweet Label}} \\ \cline{3-6} 
% \multicolumn{1}{l}{} &  & \cellcolor[HTML]{EFEFEF}Racism & \cellcolor[HTML]{EFEFEF}Sexism & \cellcolor[HTML]{EFEFEF}Both & \cellcolor[HTML]{EFEFEF}Neither \\ \hline
% \multicolumn{1}{|c|}{\cellcolor[HTML]{EFEFEF}} & \cellcolor[HTML]{EFEFEF}Racism & 16 & 0 & 0 & 1 \\ \cline{2-6} 
% \multicolumn{1}{|c|}{\cellcolor[HTML]{EFEFEF}} & \cellcolor[HTML]{EFEFEF}Sexism & 9 & 662 & 10 & 222 \\ \cline{2-6} 
% \multicolumn{1}{|c|}{\cellcolor[HTML]{EFEFEF}} & \cellcolor[HTML]{EFEFEF}Both & 0 & 4 & 0 & 1 \\ \cline{2-6} 
% \multicolumn{1}{|c|}{\multirow{-4}{*}{\cellcolor[HTML]{EFEFEF}\textbf{\begin{tabular}[c]{@{}c@{}}Similar \\Tweet \\ Label\end{tabular}}}} & \cellcolor[HTML]{EFEFEF}Neither & 26 & 754 & 5 & 218 \\ \hline
% \end{tabular}
% \end{table}
%
\begin{table}[h]
\centering
\caption{Annotation Inconsistency matrix for DT}
\label{tab:dt_matrix}
\begin{tabular}{ccccc}
\cmidrule[\heavyrulewidth]{3-5}
\multicolumn{1}{l}{} & \multicolumn{1}{l}{} & \multicolumn{3}{c}{\cellcolor[HTML]{EFEFEF}\textbf{Contentious Tweet Label}} \\ \cmidrule{3-5} 
\multicolumn{1}{l}{} &  & \cellcolor[HTML]{EFEFEF}Offensive & \cellcolor[HTML]{EFEFEF}Hate & \cellcolor[HTML]{EFEFEF}Neither \\ \midrule
\multicolumn{1}{c}{\cellcolor[HTML]{EFEFEF}} & \cellcolor[HTML]{EFEFEF}Offensive & 282 & 760 & 282 \\ %\cline{2-5} 
\multicolumn{1}{c}{\cellcolor[HTML]{EFEFEF}} & \cellcolor[HTML]{EFEFEF}Hate & 84 & 133 & 16 \\ %\cline{2-5} 
\multicolumn{1}{c}{\multirow{-3}{*}{\cellcolor[HTML]{EFEFEF}\textbf{\begin{tabular}[c]{@{}c@{}}Similar \\Tweet \\ Label\end{tabular}}}} & \cellcolor[HTML]{EFEFEF}Neither & 105 & 41 & 74 \\ \bottomrule
\end{tabular}
\end{table}
From Table \ref{tab:wz_matrix}, we observe that 662 sexism contentious tweets have their most similar tweets sharing the same label, while 745 of the sexism contentious tweets have their most similar tweets labeled as normal tweets (i.e., neither). 
\begin{table}[h!]
%\centering
\caption{Annotation Inconsistency matrix for FOUNTA}
\label{tab:founta_matrix}
\begin{tabular}{cccccc}
\cmidrule[\heavyrulewidth]{3-6}
\multicolumn{1}{l}{} & \multicolumn{1}{l}{} & \multicolumn{4}{c}{\cellcolor[HTML]{EFEFEF}\textbf{Contentious Tweet Label}} \\ \cmidrule{3-6} 
\multicolumn{1}{l}{} &  & \cellcolor[HTML]{EFEFEF}Abusive & \cellcolor[HTML]{EFEFEF}Hate & \cellcolor[HTML]{EFEFEF}Spam & \cellcolor[HTML]{EFEFEF}Normal \\ \midrule
\multicolumn{1}{c}{\cellcolor[HTML]{EFEFEF}} & \cellcolor[HTML]{EFEFEF}Abusive & 491 & 1547 & 736 & 1062 \\ 
\multicolumn{1}{c}{\cellcolor[HTML]{EFEFEF}} & \cellcolor[HTML]{EFEFEF}Hate & 347 & 370 & 93 & 192 \\
\multicolumn{1}{c}{\cellcolor[HTML]{EFEFEF}} & \cellcolor[HTML]{EFEFEF}Spam & 109 & 62 & 790 & 1024 \\ 
\multicolumn{1}{c}{\multirow{-4}{*}{\cellcolor[HTML]{EFEFEF}\textbf{\begin{tabular}[c]{@{}c@{}}Similar \\Tweet \\ Label\end{tabular}}}} & \cellcolor[HTML]{EFEFEF}Normal & 758 & 1133 & 3170 & 915 \\ \bottomrule
\end{tabular}
\end{table}
This suggests that there could be inconsistencies in the annotation of sexism tweets as two similar tweets may have different labels, one labeled as sexism while another as normal. Similar observations are made in other class labels, although the inconsistency in sexism tweet annotation is observed to be the highest. Similar observations are also made for the \textbf{DT} dataset in Table \ref{tab:dt_matrix}. A majority of the contentious hate tweets have their most similar tweets labeled as offensive. This is unsurprising as even for human annotators it is often difficult to differentiate hateful tweets from offensive ones \citep{DavidsonWMW17}. Nevertheless, such challenges in annotation also highlight the difficulty in the hate speech detection task.

Comparing the annotation inconsistency matrix of \textbf{FOUNTA} against the other two datasets, we noted that there could be significantly more annotation inconsistencies in the \textbf{FOUNTA} dataset. As shown in Table \ref{tab:founta_matrix}, there is a high amount of annotation inconsistencies observed in all labels. For instance, we observed that 758 contentious abusive tweets have their most similar tweets labeled as normal, and a significant number of contentious hate tweets have their most similar tweets labeled as abusive or normal. We further verify the annotation inconsistencies in \textbf{FOUNTA} dataset by retrieving some samples of the \textbf{FOUNTA} tweets. 
%
%
% \begin{table}[h!]
% \centering
%   \caption{Examples of tweets from \textbf{FOUNTA} dataset. C\textit{x} denotes the contentious tweet and S\textit{x} denotes the corresponding most similar tweet.}
%   \label{tab:example}
%   \begin{tabular}{c|p{14em}|c}
%     \hline
%     \textbf{id} & \textbf{tweet} & \textbf{Label}\\\hline\hline
%     C1 & RT:[USER\_1] How about we f**king hire trans boys to play trans boys & hate\\\hline
%     S1 & RT:[USER\_1] How about we f**king hire trans boys to play trans boys & normal\\\hline\hline
%     C2 & RT:[USER\_2] I wish I wasn't so annoying like I even piss myself off & normal\\\hline
%     S2 & RT:[USER\_2] I wish I wasn't so annoying like I even piss myself off & abusive\\\hline\hline
%     C3 & RT:[USER\_3] f**king faggot & hate\\\hline
%     S3 & [USER\_4] f**king faggot & abusive\\\hline\hline
% \end{tabular}%}
% \end{table}
%
Table \ref{tbl:example} shows three examples of the \textbf{FOUNTA} contentious tweets and their most similar tweets. Surprisingly, we notice that the most similar tweets retrieved for contentious tweets C1 and C2 are retweets, and the retweets are annotated with different class labels. This exposes an issue in FOUNTA's annotation strategy. We postulated that the identical tweets (i.e., the retweets) are annotated by different human annotators, resulting in the inconsistencies. We further investigated and found that more than 10\% of the tweets are retweets, and most of which have inconsistencies in their annotation. 
\begin{table}[h!]
\centering
  \caption{Examples of tweets from \textbf{FOUNTA} dataset. C\textit{x} denotes the contentious tweet and S\textit{x} denotes the corresponding most similar tweet.}
  \label{tab:example}
  \begin{tabular}{cm{15em}l}
    \toprule
    \textbf{Id} & \textbf{Tweet} & \textbf{Label}\\\midrule
    C1 & RT:[USER\_1] How about we f**king hire trans boys to play trans boys & hate\\\hdashline
    S1 & RT:[USER\_1] How about we f**king hire trans boys to play trans boys & normal\\\hline
    C2 & RT:[USER\_2] I wish I wasn't so annoying like I even p*ss myself off & normal\\\hdashline
    S2 & RT:[USER\_2] I wish I wasn't so annoying like I even p*ss myself off & abusive\\\hline
    C3 & RT:[USER\_3] f**king faggot & hate\\\hdashline
    S3 & [USER\_4] f**king faggot & abusive\\\bottomrule
\end{tabular}%}
\end{table}

\section{Conclusion}
In this paper, we proposed an analytical framework to examine annotation consistency in online misbehavior datasets. We applied our proposed framework to analyze three popular online misbehavior datasets. Our analysis showed that annotation inconsistencies in all three datasets, illustrating the challenges in online misbehavior data collection. Specifically, in the \textbf{FOUNTA} dataset, we found a significant amount of annotation inconsistency where identical tweets are annotated with different class labels. We also provided the updated datasets\footnote{ https://gitlab.com/bottle\_shop/abusive/annotation\_framework.} with annotation inconsistency information so that researchers may perform the necessary data preprocessing in future online misbehavior studies.

%\balance
\bibliographystyle{aaai}
\bibliography{ref}

\end{document}